\documentclass[prl,twocolumn,showpacs,preprintnumbers,amsmath,amssymb]{revtex4}
\usepackage{graphicx}%
\usepackage{dcolumn} %
\usepackage{bm}      %
\newcommand{\be}{\begin{equation}}
\newcommand{\ee}{\end{equation}}
\newcommand{\bea}{\begin{eqnarray}}
\newcommand{\eea}{\end{eqnarray}}
\newcommand{\mm}{\mathrm}

\newcommand{\mc}{\mathcal}

\newcommand{\bi}{\begin{itemize}}
\newcommand{\ei}{\end{itemize}}

\begin{document}

\title{Unexpected universality in static and dynamic avalanches}
  \author{Yang Liu}
  \author{Karin A.\ Dahmen}
  \affiliation{Department of Physics, University of Illinois at
    Urbana-Champaign, Urbana, Illinois 61801, USA}
  \date{\today}
\begin{abstract}
  We find that some equilibrium systems and their non-equilibrium counterparts
  actually show the same jerky response or avalanche behavior on many scales
  in response to slowly changing external conditions. In other words, their
  static and dynamic avalanches behave statistically the same. This suggests
  that their critical properties are much more generally applicable than
  previously assumed. In this case, systems far from equilibrium may be used
  to predict equilibrium critical behavior, and vice versa.
\end{abstract}
  \maketitle

  Avalanche behavior in diverse dynamical systems has been extensively studied
  in the past
  decade~\cite{Sethna-Nature-01,Durin-Book-05,Vives-PRL-94,Nori-PRL-95}.
  In those systems, there are often a large number of metastable states. When
  pushed by an external driving field, those systems shift from one metastable
  state to another, responding with collective behavior in the form of avalanches. A
  dynamic avalanche is just the rearrangement of the system configuration,
  which connects two different metastable states at two slightly different
  external fields. In experiments, avalanches are often associated with
  crackling noise as measured in acoustic emission and Barkhausen noise
  experiments~\cite{Vives-PRL-94, Durin-Book-05}. So far, avalanche behavior
  in equilibrium systems, i.e. static ``avalanche'', has rarely been studied
  due to computational complexity. With a static ``avalanche'' we refer
  to a configuration rearrangement connecting two different neighboring ground
  states at two slightly different external fields.

  Generally, equilibrium systems are believed to be completely different from
  non-equilibrium ones simply because the underlying physics is so
  different. A natural question arises: Do static and dynamic avalanches have
  the same critical behavior? Answering this basic question would be crucial
  to understand whether there are any possible deep connections between
  equilibrium systems and their non-equilibrium counterparts.

  In this letter, we show compelling evidence that static and dynamic
  avalanches have the same critical behavior in the zero-temperature
  random-field Ising model (zt-RFIM). This particular model is chosen for
  two reasons.  First, there is a related highly controversial question in
  this model, i.e. whether the equilibrium and non-equilibrium
  disorder-induced phase transitions belong to the same universality
  class. Second, both static and dynamic avalanches can be clearly identified
  and calculated within this model. We find that all tested universal scaling
  functions and corresponding critical exponents coincide for static and
  dynamic avalanches. Our findings indicate that generally equilibrium systems
  and their non-equilibrium counterparts may have deep
  connections~\cite{Maritan-PRL-94,PerezReche-PRB-04}. 
  As a prototypical model for disordered magnets, the RFIM has been
  intensely studied~\cite{Nattermann-Chapter-98}. Its Hamiltonian is given
  by $ {\cal H} = - \! J \sum_{{<}i,j{>}} \, s_i s_j - \sum_i \, (H + h_i)
  \,  s_i $ where the Ising spins $s_i = \pm 1$ sit on a $d$-dimensional
  hypercubic lattice with periodic boundary conditions. The spins interact
  ferromagnetically with their nearest neighbors with strength $J$  and
  experience a uniform external field $H$ and a quenched local random field
  $h_i$. Usually, the local fields are chosen from a Gaussian distribution
  $\rho(h)$ with mean 0 and standard deviation $R$. $R$ is called the disorder
  parameter. In equilibrium, it is generally believed that in $d>2$ the
  transition between the ordered (ferromagnetic) and disordered (paramagnetic)
  phases is continuous and controlled by a stable zero-temperature fixed
  point~\cite{Fisher-PRL-86}. Therefore, one can set temperature $T=0$ and
  tune disorder $R$ to study the equilibrium disorder-induced phase transition
  (DIPT) undergone by the ground-state
  properties. In non-equilibrium, the
  zt-RFIM has been very successful in explaining dynamic avalanches and
  crackling noise observed in magnets~\cite{Sethna-PRL-93, Dahmen-PRB-96,
  Perkovic-PRB-99}. The key result is that there is a non-equilibrium DIPT,
  associated with the hysteretic behavior. Based on the similarities of some 
  critical exponents, it has been conjectured that the equilibrium and
  non-equilibrium DIPT may belong to the same universality class. But this has
  been highly controversial due to the lack of compelling
  evidence~\cite{Maritan-PRL-94,
  Sethna-PRL-94,PerezReche-PRB-04,Colaiori-PRL-04}. Comparing the critical
  behavior of static and dynamic avalanches is important and necessary to
  answer this question.

  An avalanche in the RFIM refers to the flip of neighboring spins during
  the magnetization process, corresponding to a jump in the magnetization
  curve $M(H)$. To identify avalanches, we increase the external field $H$
  from $-\infty$ to $\infty$ adiabatically slowly, i.e. $H$ is kept constant
  during the propagation of an avalanche. Then, a static (dynamic) avalanche 
  connects two \emph{nearest} ground (metastable) states along the equilibrium
  (non-equilibrium) $M(H)$ curve at zero-temperature. In equilibrium, the ground
  state problem of the RFIM can be mapped onto the min-cut/max-flow problem of
  a network and solved via the so-called push-relabel algorithm~\cite{Middleton-PRB-02,Dukovski-PRB-03}. An efficient
  linear interpolation scheme is then used to find steps by narrowing down the
  $H$ range where static avalanches occur~\cite{Frontera-JCP-00}. In non-equilibrium,
  we use the single-spin-flip dynamics to calculate the metastable state: each
  spin flips deterministically when its effective local field $h^{\rm eff}_i =
  J \sum_{j} s_j + h_i + H $ changes
  sign~\cite{Sethna-PRL-93,Perkovic-PRB-99}. Due to the nearest neighbor
  interaction, a flipped spin will push a neighbor to flip, which in turn
  might push another neighbor, and so on, thereby generating an dynamic
  avalanche.

  To study whether the shape of the random field distribution would
  affect the avalanche behavior, we consider four different types of
  $\rho(h)$'s: (1) Gaussian: $\rho_\mm{G}(h)=\frac{1}{\sqrt{2\pi} R} \exp(-\frac{h^2}{2R^2})$;
  (2) Lorentzian: $\rho_\mm{L}(h)=\frac{1}{2\pi} \frac{R}{h^2+(R/2)^2}$; (3)
  parabolic: $\rho_\mm{P}(h)=\frac{R^2-h^2}{4R^3/3}$ for $h\in [-R,R]$ and 0 else; (4)
  uniform: $\rho_\mm{U}(h)=\frac{1}{2R}$ for $h\in [-R,R]$ and 0 else. In all cases,
  $\rho(h)$ is symmetric around $h=0$ and the generalized ``width'' $R$ will be
  called the disorder parameter.

  Fig.1 shows the $M(H)$ curves and corresponding avalanches occurring
  during the magnetization processes at different disorders in both
  equilibrium (d,e,f) and non-equilibrium (a,b,c). For $R<R_\mm{c}$ (a,d),
  most spins tend to flip collectively in a system spanning avalanche seen as
  a macroscopic jump in the magnetization curve. For $R>R_\mm{c}$ (c,f), spins
  tend to flip individually and result in many microscopic avalanches and a
  macroscopically smooth magnetization curve. For $R \sim R_\mm{c}$ (b,e),
  jumps (avalanches) of all sizes are seen in the magnetization
  curve. Qualitatively, we find that static and dynamic avalanches show
  similar disorder dependent behavior.

  \begin{figure}
    \includegraphics[width=0.46\textwidth]{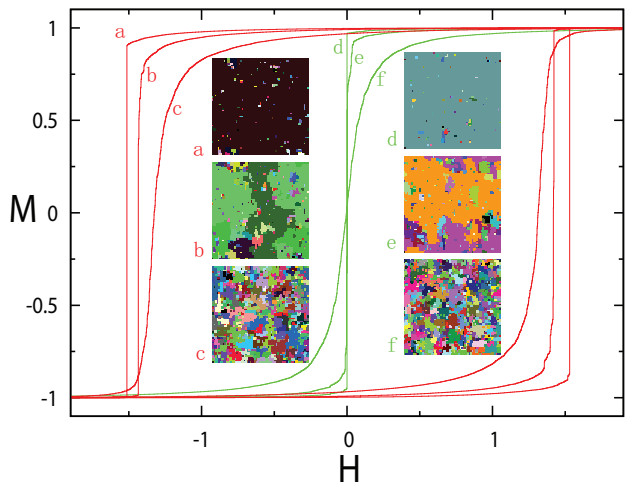}
    \caption{\label{fig:DIPT} (Color online) Disorder dependent avalanche behavior in
      RFIM. Main panel: Magnetization curves in equilibrium (green) and
      non-equilibrium (red), below, near and above the critical disorder
      $R_\mm{c}$. Insets: Cross-sections of 3D systems showing all the
      avalanches (denoted by different colors) occurring during those
      magnetization processes. The calculation is done on 3D Gaussian zt-RFIM
      with system size $64^3$. Non-equilibrium: (a) $R=2.0$, (b) $R=2.224$,
      (c) $R=2.6$. Equilibrium: (d) $R=2.25$, (e) $R=2.45$, (f) $R=2.9$.  Note
      that $R^\mm{neq}_\mm{c}=2.16 \pm 0.03$ and $R^\mm{eq}_\mm{c}=2.28 \pm
      0.01$ for 3D Gaussian RFIM~\cite{Perkovic-PRB-99,Hartmann-PRB-01}.}
  \end{figure}
  To quantitatively study the similarity of static and dynamic avalanches, we
  first study the avalanche size distribution integrated over the external
  field~\cite{Perkovic-PRB-99}. Near the critical disorder $R_\mm{c}$, its scaling form can be
  written as
  \be D_\mm{int}(S,R) \sim
  S^{-(\tau+\sigma\beta\delta)} \ \bar{\mc{D}}_{\pm}^{\mm{int}}
  (S^{\sigma}|r|) \label{Dint-SR} \ee where $S$ is the avalanche size, i.e. the number of
  spins participating in an avalanche, $\pm$ refers to the sign of the reduced
  disorder $r=(R_\mm{c}-R)/R$, $\sigma$ gives the scaling of the largest
  avalanche size $S_\mm{max} \sim |r|^{-1/\sigma}$, $\beta$ and $\delta$ give
  the singularities of $M(H)$ near the critical point
  $(H_\mm{c},R_\mm{c})$. Here, the critical field $H_\mm{c}$ is defined to be
  the field where the slope of $M(H)$ goes to $\infty$. In non-equilibrium,
  $D_\mm{int}(S,R)$ for Gaussian $\rho(h)$ has been studied extensively. The
  critical exponents $(\tau+\sigma\beta\delta)=2.03 \pm 0.03$,
  $\sigma=0.24\pm0.02$ and the universal scaling function
  $\bar{\mc{D}}_{-}^{\mm{int}} (X) = e^{-0.789 X^{1/\sigma}} (0.021 + 0.002 X
  + 0.531 X^2 - 0.266 X^3 + 0.261X^4 )$ were obtained from scaling collapses
  of $D_\mm{int}(S,R)$ at different disorders~\cite{Perkovic-PRB-99}.

  Fig.2(a) shows that for Gaussian, Lorentzian and parabolic $\rho(h)$'s, and
  for both static and dynamic avalanches at different disorders, with the
  same pair of critical exponents: $(\tau+\sigma\beta\delta)=2.03$ and
  $\sigma=0.24$, 24 $D_\mm{int}(S,R)$ curves collapse onto a single one. The
  universality that the three different $\rho(h)$'s show the same avalanche
  behavior is not a surprise at all. A renormalization group (RG) analysis has
  shown that, at least in non-equilibrium, what matters is just $\rho''(0)$,
  i.e. the second derivative of $\rho(h)$ at $h=0$~\cite{Dahmen-PRB-96}. It is
  easy to check that Gaussian, Lorentzian and parabolic $\rho(h)$'s all have
  $\rho''(0) \sim -R^3$. Therefore their universal behaviors agree, as
  expected. Fig.2(b) shows that for uniform $\rho(h)$, for both static and
  dynamic avalanches at different disorders, eight $D_\mm{int}(S,R)$ curves
  collapse onto a single one, with critical exponents:
  $(\tau+\sigma\beta\delta)=2.08\pm0.02$ and $\sigma=0.52\pm0.03$. Note that:
  (1) The critical exponents, especially $\sigma$, are significantly different
  from those of the above three kinds of $\rho(h)$'s. (2) The scaling function
  has a significantly different shape from that observed in Fig.2(a). These
  two findings are consistent with the RG analysis mentioned above because for
  a uniform $\rho(h)$, $\rho''(0)=0$ is independent of $R$ and is
  qualitatively different from the other three distributions.

  The most surprising result about Fig.2 is that the critical exponents and
  scaling functions for static and dynamic avalanches match for any
  $\rho(h)$. This strongly indicates that the equilibrium and non-equilibrium
  RFIM behave the same near their corresponding critical points.

  \begin{figure}
     \scalebox{1}[0.8]{\includegraphics[width=0.46\textwidth]{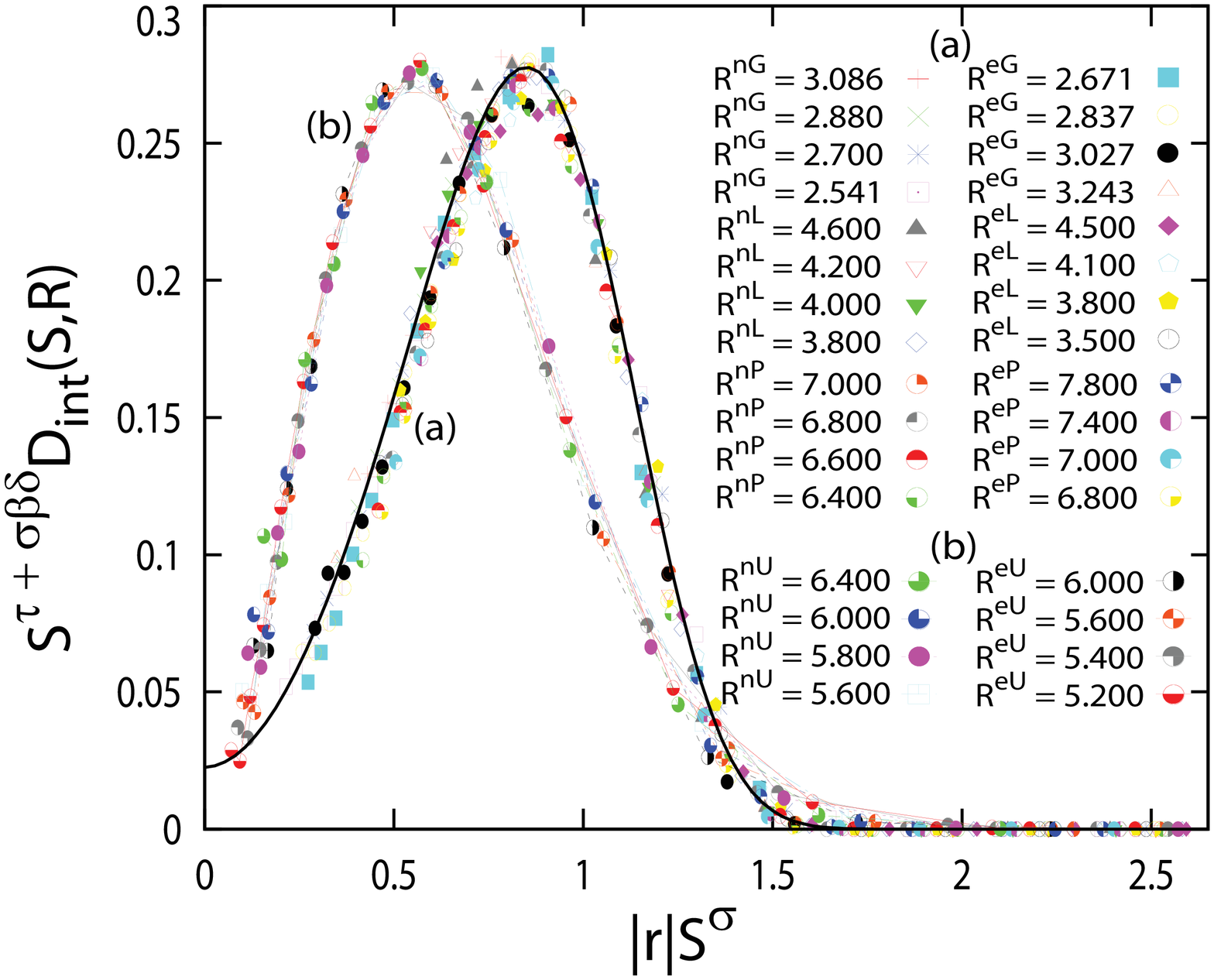}}
    \caption{\label{fig:DintSall}  (Color online) Scaling functions of integrated
      avalanche size distribution. The original $D_\mm{int}(S,R)$ curves for
      static and dynamic avalanches at different disorders and different
      $\rho(h)$'s are calculated in 3D with system size $64^3$ and are
      averaged up to 100 initial random-field configurations. In the legend,
      the subscripts stand for equilibrium (e) or non-equilibrium (n) and the
      type of $\rho(h)$: Gaussian (G), Lorentzian (L), parabolic (P), and
      uniform (U). (a) For Gaussian, Lorentzian and parabolic $\rho(h)$'s,
      using critical exponents $(\tau+\sigma\beta\delta)=2.03$ and
      $\sigma=0.24$, 24 curves collapse onto each other, up to non-universal
      critical disorders ($R^\mm{nG}_\mm{c}=2.16$, $R^\mm{eG}_\mm{c}=2.29$;
      $R^\mm{nL}_\mm{c}=1.92$, $R^\mm{eL}_\mm{c}=2.08$;
      $R^\mm{nP}_\mm{c}=4.84$, $R^\mm{eP}_\mm{c}=5.0$) and overall scale
      factors.  The thick black curve through the collapse is
      the universal scaling function $\bar{\mc{D}}_{-}^{\mm{int}} (X)$ of
      non-equilibrium Gaussian RFIM~\cite{Perkovic-PRB-99}. (b) For uniform
      $\rho(h)$, eight curves collapse onto each other, up to non-universal
      critical disorders ($R^\mm{nU}_\mm{c}=4.64$, $R^\mm{eU}_\mm{c}=4.46$)
      and overall scale factors. The collapse yields critical disorders 
      $(\tau+\sigma\beta\delta) = 2.08 \pm 0.02$ and $\sigma=0.52 \pm 0.03$.
      Both the critical exponents and the scaling function are different from
      those of the Gaussian $\rho(h)$.}
  \end{figure}

  To check whether this is just a coincidence, we make another
  independent test by studying the avalanche correlation function,  which
  measures the probability that a distance $x$ between any two flipping spins
  occurs in the same avalanche~\cite{Perkovic-PRB-99}. Near the critical
  disorder $R_\mm{c}$, the scaling form of the avalanche correlation function
  integrated over $H$ can be written as
  \be
  G_\mm{int}(x,R) \sim \frac{1}{x^{d+\beta/\nu}} \bar{\mc{G}}_\pm(x|r|^\nu)
  \ee with $\nu$ the correlation length exponent and $d$ the
  dimension~\cite{Dahmen-PRB-96}. In non-equilibrium, the quantity
  $G_\mm{int}(x,R)$ for the Gaussian $\rho(h)$ has been studied
  extensively, where $d+\beta/\nu=3.07\pm 0.30$ and $\nu=1.37\pm 0.18$ were
  obtained from scaling collapses of $G_\mm{int}(x,R)$ at different
  disorders~\cite{Perkovic-PRB-99}. Here, in Fig.3(a), we show that for Gaussian,
  Lorentzian and parabolic $\rho(h)$'s, and for both static and dynamic avalanches
  at different disorders, with the same pair of critical exponents:
  $d+\beta/\nu=3.07$ and $\nu=1.37$, 24 $G_\mm{int}(x,R)$ curves collapse onto
  a single one. Fig.3(b) shows that for uniform $\rho(h)$ and for both
  static and dynamic avalanches at different disorders, eight
  $G_\mm{int}(x,R)$ curves collapse onto a single one, with critical  
  exponents: $d+\beta/\nu=3.0 \pm 0.3$ and $\nu=0.8 \pm 0.2$. On one hand, both the
  critical exponents and the scaling function for the uniform $\rho(h)$ are different
  from those of the Gaussian $\rho(h)$. On the other hand, both the critical
  exponents and scaling functions for static and dynamic avalanches match for
  any $\rho(h)$. These results are completely consistent with what we found in
  avalanche size distributions.

  \begin{figure}
    \scalebox{1}[0.8]{\includegraphics[width=0.46\textwidth]{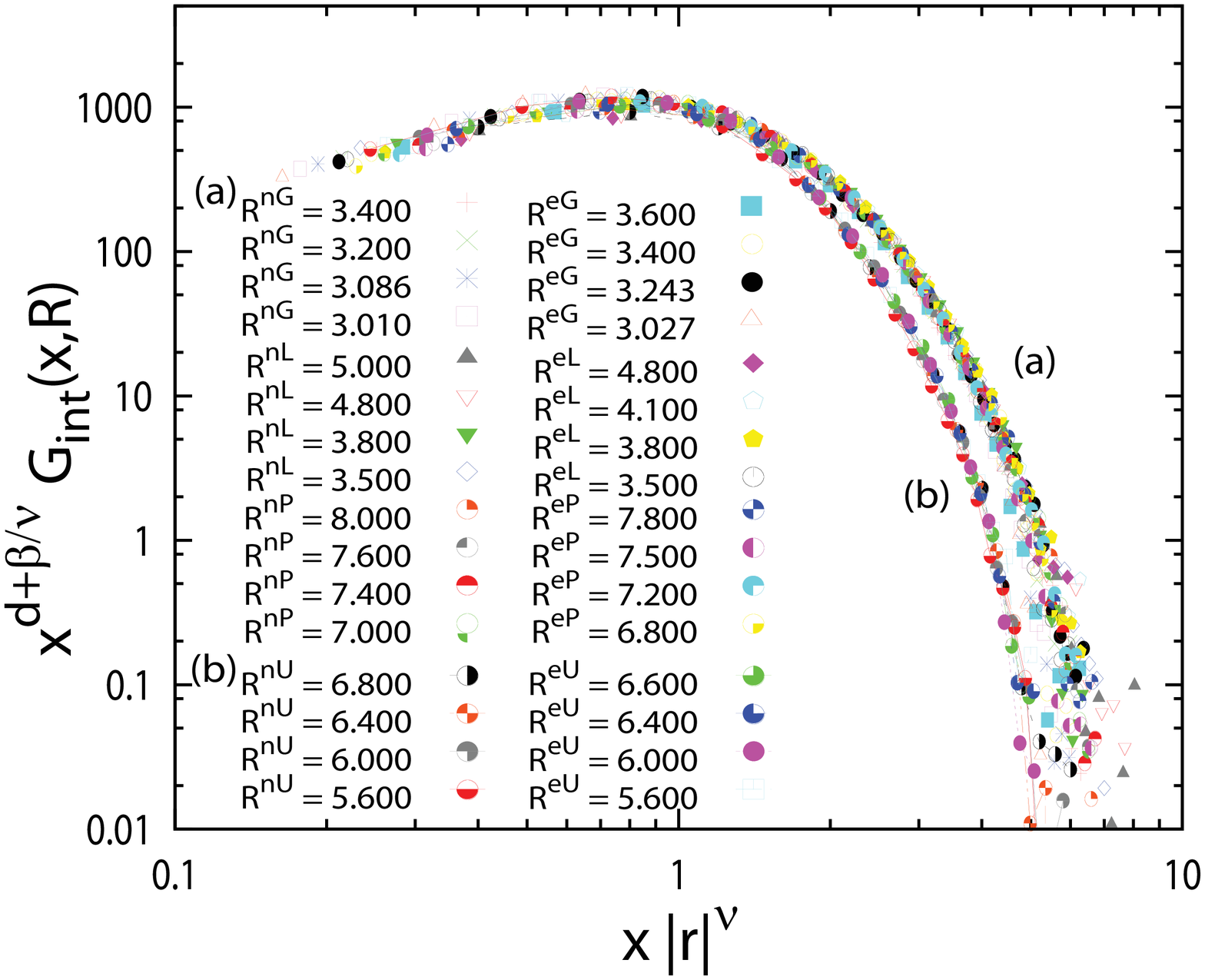}}
    \caption{\label{fig:Gxall}  (Color online) Scaling functions of integrated
      avalanche correlation function. The original $G_\mm{int}(x,R)$ curves
      for static and dynamic avalanches at different disorders and different
      $\rho(h)$'s are calculated in 3D with system size $64^3$ and are
      averaged up to 100 initial random-field configurations. (a) For Gaussian,
      Lorentzian and parabolic $\rho(h)$'s, 24 curves collapse onto each
      other, using $d+\beta/\nu=3.07$ and
      $\nu=1.37$~\cite{Perkovic-PRB-99}. (b) For uniform $\rho(h)$, eight
      curves collapse onto each other with $d+\beta/\nu=3.0 \pm 0.3$ and
      $\nu=0.8 \pm 0.2$. Note that those collapses are up to the same critical
      disorders as used in Fig.2.} 
  \end{figure}

  Moreover, in a separate work, for the Gaussian $\rho(h)$, we have shown that
  static and dynamic avalanches have surprisingly similar spatial structures
  with the same fractal dimensions, anisotropy measures, and associated
  universal scaling functions~\cite{Liu-arXiv-06}.  More interestingly, we notice that the
  equilibrium and non-equilibrium DIPTs themselves show surprising similarity:
  (1) They share the same \emph{no-passing rule}: at $T=0$, flipped spins can never flip
  back as the magnetic field $H$ is swept monotonically~\cite{Liu-PRE-07,
  Sethna-PRL-93}. (2) In mean field theory, they have the same thermodynamic
  critical exponents~\cite{Schneider-PRB-77, Sethna-PRL-93}, the same
  avalanche critical exponents~\cite{Liu-arXiv-06}, and the same exponent
  relations~\cite{Dahmen-PRB-96}. (3) RG calculations show that the
  $6-\epsilon$ expansion for the non-equilibrium critical exponents maps to
  all orders in $\epsilon$ onto the controversial equilibrium ones: The temperature
  dependence is irrelevant in the equilibrium RFIM and the time dependence is
  irrelevant in the zero-temperature non-equilibrium RFIM, leaving us with the
  same starting point for the calculation in both
  cases~\cite{Dahmen-PRB-96}. All these evidences in favor of universality
  corroborate our findings here.

  To discuss the effect of dynamics on the critical behavior of avalanches, a
  general $k$-spin-flip dynamics (with $k=1,2,\cdots,\infty$) has been
  introduced~\cite{PerezReche-PRB-04}. It is defined such that all the states
  connected by avalanches are $k$-spin-flip metastable states whose energy
  cannot be lowered by the flip of any subset of $1,2,\cdots,k$
  spins~\cite{Newman-PRE-99, PerezReche-PRB-04}. The case $k=1$ just corresponds to the
  single-spin-flip dynamics used in our non-equilibrium calculations. The case
  $k=\infty$ corresponds to the ground state evolution dynamics in our
  equilibrium calculations. It has been found that the change of dynamics from
  $k=1$ to $k=2$ will not alter the critical behavior of the dynamic
  avalanches~\cite{Vives-PRB-05}. Together with our finding, i.e. $k=1$ and
  $k=\infty$ give the same avalanche behavior, we suggest that avalanches
  associated with the whole series of $k$-spin-flip dynamics (with
  $k=1,2,\cdots \infty$) would have the same critical behavior.
  The $k$-spin-flip dynamics is quite general but it definitely cannot
  encompass all kinds of dynamics, e.g. the demagnetization dynamics
  associated with the demagnetization curve, which is obtained by applying an
  oscillating external field with very slowly decreasing amplitude. %
  Previous studies for Gaussian $\rho(h)$ showed that there are two very
  interesting results. First, the avalanches associated with the
  demagnetization curve are found (within numerical error bars) to display the
  same critical exponents and scaling functions as the avalanches associated
  with the saturation hysteresis loop (with $k=1$)~\cite{Carpenter-PRB-03}. Second, the
  demagnetized state and ground state show similarity near their corresponding
  critical disorders: the critical exponents and scaling function associated
  with the $M(R)$ curve coincide~\cite{Colaiori-PRL-04}.

  \begin{figure}
     \scalebox{1}[0.8]{\includegraphics[width=0.46\textwidth]{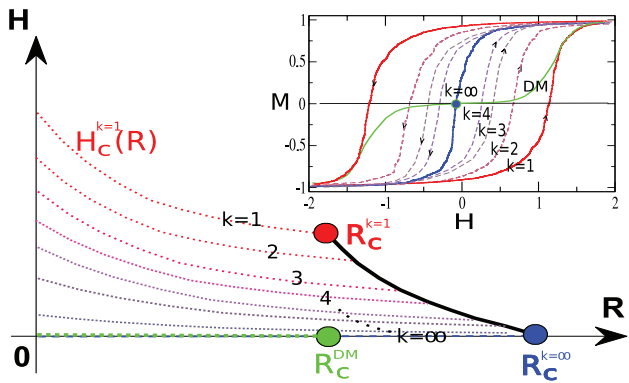}}
    \caption{\label{fig:pd}  (Color online) Schematic phase diagram of zt-RFIM with
    $k$-spin-flip dynamics and demagnetization dynamics. Dashed lines stand
    for the first-order phase transitions occurring at the critical field
    $H_\mm{c}$. Note that for both demagnetization dynamics and ground state
    evolution ($k=\infty$), $H_\mm{c}=0$ due to symmetry. (Inset) Schematic
    $M(H)$ curves associated with different dynamics.}
  \end{figure}

  Considering all the findings, we suggest that all the different dynamics
  yield the same scaling behavior of avalanches --- an unexpected
  universality, see Fig.4. It would be very interesting to numerically test this
  universality in other disordered systems, especially for those systems with
  frustrations where the no-passing rule is broken. We suspect that a
  necessary condition for equilibrium and non-equilibrium critical behavior to
  scale in the same way is that the scaling behavior is dominated in both
  cases by a zero-temperature fixed point~\cite{Fisher-PC-08}. For example,
  for the random-bond Ising model, which has a non-trivial finite-temperature
  fixed point~\cite{Hukushima-JPSJ-00}, the equilibrium and non-equilibrium
  critical behavior are different~\cite{Dahmen-PRB-96}.

  We thank James P. Sethna, A. Alan Middleton, Yoshitsugu Oono, Daniel
  S. Fisher, Gil Refael, Andrei A. Fedorenko, Charles M. Newman and Daniel
  L. Stein for valuable discussions. We acknowledge the support of NSF Grant
  No. DMR 03-14279 and NSF Grant No. DMR 03-25939 ITR (Materials Computation
  Center). 

  \bibliography{RFIM}%
\end{document}